\documentclass{iopart}

\usepackage{pslatex}
\usepackage{iopams}
\usepackage[dvips]{graphicx}
\usepackage{color}

\begin{document}

\title[Calibration of the LIGO displacement actuators via laser frequency modulation]{Calibration of the LIGO displacement actuators via laser frequency modulation}

\author{E~Goetz$^1$ and R~L~Savage~Jr$^2$}
\address{$^1$ University of Michigan, Ann Arbor, MI 48109, USA}
\address{$^2$ LIGO Hanford Observatory, Richland, WA 99352, USA}

\eads{\mailto{savage\_r@ligo-wa.caltech.edu} and \mailto{egoetz@umich.edu}}

\begin{abstract}
We present a frequency modulation technique for calibration of the displacement actuators of the LIGO 4-km-long interferometric gravitational-wave detectors. With the interferometer locked in a single-arm configuration, we modulate the frequency of the laser light, creating an effective length variation that we calibrate by measuring the amplitude of the frequency modulation. By simultaneously driving the voice coil actuators that control the length of the arm cavity, we calibrate the voice coil actuation coefficient with an estimated $1\sigma$ uncertainty of less than one percent. This technique enables a force-free, single-step actuator calibration using a displacement fiducial that is fundamentally different from those employed in other calibration methods.
\end{abstract}

\pacs{04.80.Nn, 06.30.Bp, 07.60.Ly, 42.60.Da, 95.55.Ym}

\section{Introduction}\label{Intro}
The LIGO (Laser Interferometer Gravitational-wave Observatory) interferometers are sensitive to differential length variations with amplitude spectral densities on the order of $10^{-19}$~m$/\sqrt{\textrm{Hz}}$. To enhance the sensitivity to gravitational waves, the interferometer arms incorporate 4-km-long optical cavities, Fabry-Perot resonators, formed by 10-kg input and end mirrors, or test masses. They are suspended as pendulums for isolation from seismic motion and to make them ``freely-falling'' test masses in the horizontal plane. To hold these cavities on resonance with the laser light, the position of the end test masses (ETMs) are controlled via voice coil actuators utilizing rare-earth magnets bonded to the back surfaces of the mirrors.

The distortions in space-time induced by passing gravitational waves will be sensed by the interferometers as differential arm length fluctuations and therefore suppressed by the differential arm length (DARM) control loop via the ETM voice coil actuators. Reconstructing the waveforms of the gravitational wave disturbances thus requires deconvolving the closed-loop response of the DARM servo. This requires characterization and calibration of the ETM voice coil actuators. Toward this end, several voice coil calibration procedures have been developed within LIGO~\cite{OcalPaper,LIGOPcal}. Detector calibration is also a significant effort at other gravitational wave observatories~\cite{GEOCal,VirgoCal}.

The method traditionally employed by LIGO, the so-called {\it free-swinging Michelson} method, uses the wavelength of the laser light as a fiducial~\cite{S5paper}. It begins with a series of measurements to deduce the actuation coefficient of the input test mass (ITM). The actuation coefficient for the ETM is then determined from the ITM coefficient via transfer function measurements made in a single-arm lock configuration. This technique has the disadvantage that it requires combining several measurements, some made in different interferometer configurations, and sequential calibration of two sets of voice coil actuators. This complexity increases statistical uncertainties and the potential for systematic errors.

Another calibration method, the {\it photon calibrator}, that induces a calibrated length variation via the reflection of a power-modulated auxiliary laser beam from the ETM surface is also being employed by LIGO and other interferometric gravitational wave detectors~\cite{LIGOPcal,GEOPcal,VirgoPcal}. This method is applied on-line in the ``science mode'' configuration used for gravitational wave searches and induces displacements similar to those expected to be caused by gravitational waves. It relies on absolute calibration of the modulated laser power reflecting from the ETM.

In this article, we describe a fundamentally different method that is based on frequency modulation of the laser light. For this technique, the interferometer is operated in a single-arm configuration. The frequency of the laser light is sinusoidally modulated, and this modulation is interpreted by the sensor of the arm cavity locking servo as a length modulation, providing a fiducial for ETM voice coil actuator calibration. This method has been used both to calibrate the LIGO actuators and to investigate systematic errors associated with other calibration methods~\cite{CalCompare}.

\begin{figure}
   \begin{flushright}
   	\includegraphics[width=0.8\textwidth]{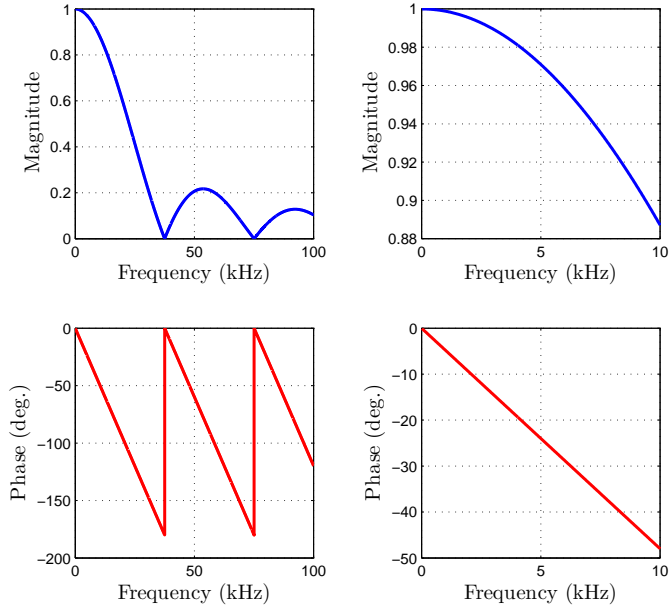}
   \end{flushright}
   \caption{Magnitude and phase of the frequency-to-length transfer function, $C(f)$, for the LIGO 4-km-long arm cavities over a 100~kHz span (left) and over a 10~kHz span (right).}
   \label{fig:CofF}
\end{figure}

For a resonant Fabry-Perot cavity, frequency variations and length variations are related by the dynamic resonance condition~\cite{DynamResonance} which is given by
\begin{equation}\label{eq:dynresonance}
   C(f) \frac{\Delta {\nu}(f)}{\nu} = -  \frac{\Delta L(f)}{L}.
\end{equation}
Here $f$ is the frequency of the variations, $\nu$ is the laser frequency, $L$ is the cavity length, $\Delta {\nu}(f)$ and $\Delta {L}(f)$ are the amplitudes of the sinusoidal variations, and $C(f)$ is the normalized frequency-to-length transfer function given by
\begin{equation}\label{eq:CofF}
C(f) = \frac{1 - e^{-4 i \pi f T}}{4 i \pi f T},
 \end{equation}
where $T=L/c$ is the light transit time in the cavity. $C(f)$ for the LIGO arm cavities is plotted in figure~\ref{fig:CofF}. A calibrated frequency modulation, $\Delta\nu$, thus results in a calibrated length modulation, $\Delta L$, that is the fiducial for the frequency modulation calibration method.

A key advantage of this technique is that it does not exert localized forces on the test masses in addition to the forces exerted by the voice coil actuators. Such additional calibration forces can cause elastic deformation of the test masses and can be a dominant source of systematic errors for other calibration methods~\cite{LIGOPcal,HildEffect}. The frequency modulation method could thus enable investigation of the elastic deformation induced by the voice coil actuator forces~\cite{TMdeform}.

\section{Calibration of the displacement actuator}\label{Alcal}
\begin{figure}
   \begin{flushright}
      \includegraphics[width=1.0\textwidth]{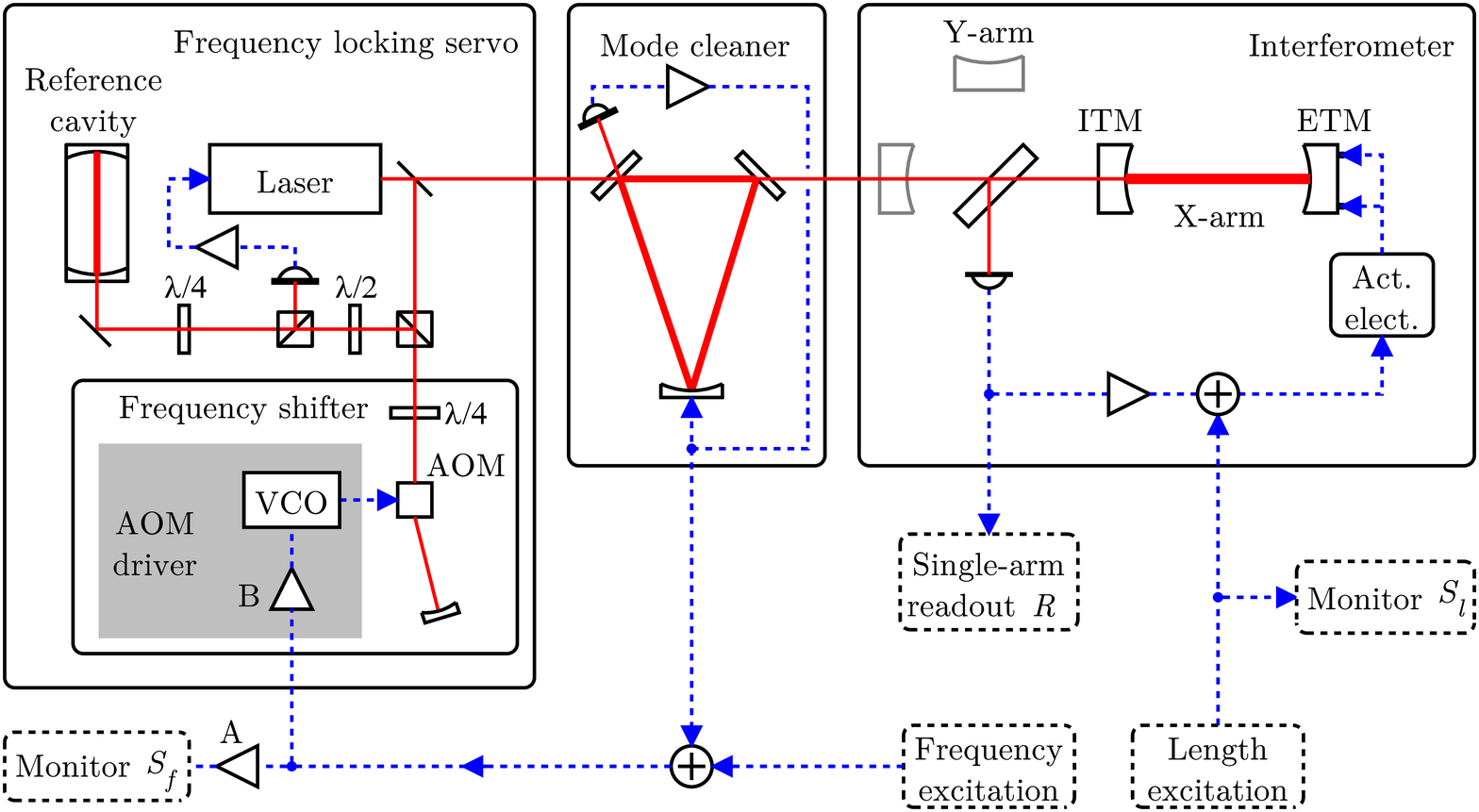}
   \end{flushright}
   \caption{Schematic of the experimental setup used to calibrate the ETM voice coil actuators using the frequency modulation technique. The laser frequency is locked to a resonance of the reference cavity. Driving the VCO input injects a frequency modulation into the frequency locking servo loop via the double-passed AOM. The frequency servo acts on the laser frequency to cancel the injected modulation, thus imposing the inverse of the modulation on the laser light directed to the mode cleaner. The mode cleaner filters the frequency modulated light which then impinges on the arm of the interferometer. The arm length is held on a resonance by the voice coil actuators that control the position of the ETM.}
   \label{FreqShifter}
\end{figure}

For a Michelson interferometer, laser frequency modulation is a ``common mode'' excitation. Therefore, to calibrate the voice coil actuators for a single test mass, the interferometer is operated in a single-arm lock configuration, as shown schematically in figure~\ref{FreqShifter}. Assuming that the coil driver electronics transfer functions are flat over the frequency range of our measurements, we expect the voice coil actuators to deliver longitudinal forces that are independent of the drive frequency. Our measurements are made at frequencies well above the 0.75 Hz pendulum resonance frequencies of the suspended test masses, so we expect them to behave as free masses with displacements that are 180 degrees out of phase with the forces from the voice coils and decreasing with the square of the drive frequency.

To calibrate a voice coil actuation coefficient at a given frequency, we lock the particular single arm of the interferometer and simultaneously drive both the laser frequency actuator and the voice coil actuators with sinusoids at frequencies separated by 0.1 Hz. Simultaneous excitation minimizes the influence of temporal variations in interferometer and control loop parameters such as optical gain changes due to alignment fluctuations. We monitor the magnitudes of the induced modulations in the arm locking servo readout signal, the frequency modulation drive signal, and the voice coil drive signal\footnote{Note that the actuator for the arm locking servo is also the voice coil actuator for the excited test mass, so the readout signal, $R$, indicates the residual length modulation sensed by the servo. The servo suppresses the frequency modulation excitation by actuating on the length of the arm via the voice coil, thus inducing a physical length variation to reduce the effective length variation.}. With the laser frequency excitation providing an independent and calibrated effective arm length variation, the ratio of the signals yields the voice coil actuation coefficient.

The laser frequency actuator modulates the laser frequency via a frequency shifter, composed of a double-passed acousto-optic modulator (AOM) and an AOM driver (see appendix), that is embedded within a laser frequency locking servo as shown in figure~\ref{FreqShifter}. The voltage-controlled oscillator (VCO) at the heart of the AOM driver operates at a nominal frequency of 80 MHz. This frequency changes in response to the AOM driver input signal. The unity gain frequency of the frequency locking servo is approximately 600 kHz and the gain at 100 kHz is more than 25 dB. Thus, for the frequencies of interest for our measurements ($<2$ kHz), changes in the laser output frequency induced by changes in the AOM driver input signal are equal and opposite to the frequency changes induced by the double-passed AOM.

Calibration of the laser frequency actuator is described in detail in the appendix. The upper plot of figure~\ref{calVCOdata} shows the results of calibrating the digitized frequency modulation monitor point signal, $S_f$, with respect to the frequency modulation of the AOM driver output and the fit to a model with a frequency-independent VCO actuation coefficient. The normalized deviation between the measured values of the monitor point calibration, $\mathcal{K}$, and the fit (meas./fit-1) is shown in the lower panel of figure~\ref{calVCOdata}. The standard error relative to the model is 0.1\%, dominated by statistical variations in the measurement of the sideband-to-carrier power ratio (see section~\ref{Errors}).
\begin{figure}
  \begin{flushright}
   \includegraphics[width=0.8\textwidth]{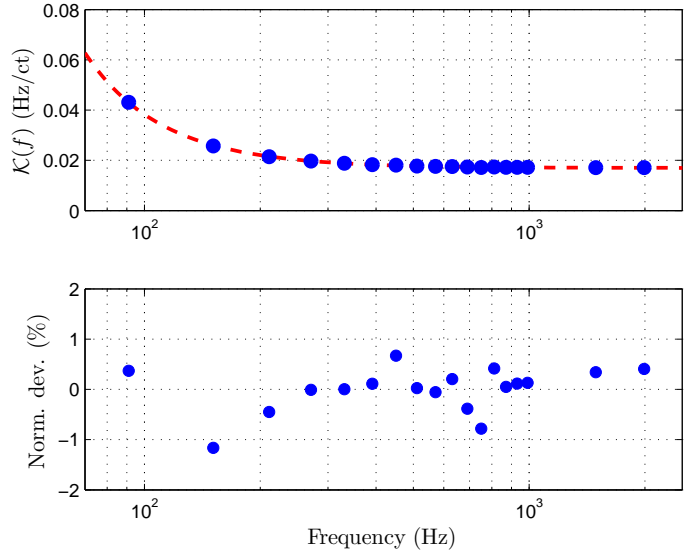}
  \end{flushright}
  \caption{Calibration function, $\mathcal{K}$, for the $S_f$ monitor point. The circles in the upper panel indicate measured values, and the dashed line is a least-squares fit assuming a frequency-independent VCO actuation coefficient, $\alpha$. The normalized deviations between the measurements and the fit are plotted in the lower panel.}
  \label{calVCOdata}
\end{figure}

As shown in figure~\ref{FreqShifter}, laser light with the frequency modulation imposed by the frequency locking servo is transmitted through a mode cleaner before impinging on the arm cavity ITM. This 12-m-long, triangular Fabry-Perot resonator has an optical storage time of approximately 35~$\mu$s that filters the laser frequency variations. To characterize the mode cleaner's passive filtering, we measure the power modulation transfer function from 10 Hz to 10 kHz using photodetectors located upstream and downstream of the mode cleaner. For modulation frequencies well below the mode cleaner cavity's free spectral range of 12.3 MHz, the response to power variations is functionally equivalent to the response to frequency variations, and can be approximated by a single real pole at frequency $f_0$~\cite{DynamResonance,MalikThesis}. Fitting the magnitude and phase of this transfer function with a single real pole yields $f_0=4.61$ kHz.

The amplitude of the frequency modulation downstream of the mode cleaner is thus given by
\begin{equation}
\Delta\nu_t(f) = \Delta\nu(f)\, H_{mc}(f) \simeq \Delta\nu(f)\left| \frac{f_0}{f_0 + i f} \right| = \frac{ \Delta\nu(f)}{\sqrt{1 + f^2 / f_0^2}}
\label{eq:dnut}
\end{equation}
where $\Delta \nu$ and $\Delta \nu_t$ are the amplitudes of the incident and transmitted sinusoidal frequency modulations and $H_{mc}$ is the frequency modulation transfer function of the mode cleaner. The servo that holds the mode cleaner on resonance actuates both the length of the mode cleaner (at low frequencies) and the frequency of the laser light (at higher frequencies) via the AOM driver (see figure~\ref{FreqShifter}). To ensure that length actuation by the mode cleaner locking servo is not changing the expected filtering function of the mode cleaner, a digital notch filter, centered at the measurement frequency, is inserted into the length control path.

Using equations~\ref{eq:dynresonance} and \ref{eq:dnut}, the effective length modulation induced by the frequency-modulated light is given by
\begin{equation}
\Delta L_f(f) \simeq \frac{-C(f)L}{\nu} \Delta \nu_t(f) = \frac{-C(f)L}{\nu} \, 2 \mathcal{K}(f) \, H_{mc}(f) \, S_f = A_f(f) \, S_f
\label{deltaLeff}
\end{equation}
where the subscript $f$ identifies terms associated with the laser frequency modulation. Here, $A_f$ is the actuation function for length variations induced by the laser frequency modulation. It converts the AOM driver input monitor signal, $S_f$, to effective arm cavity length variation. This calibrated actuation function can be used to directly calibrate the single-arm readout signal or to calibrate the ETM voice coil actuators.

Calibration of the voice coil actuators proceeds with sinusoidal excitations of the frequency and length actuators made at closely spaced frequencies in a single-arm, closed-loop configuration. The resulting length modulations appear as two peaks in the power spectrum of the single-arm readout signal. The ratio of the amplitudes of these two modulations together with monitors of the actuation amplitudes yields the ETM voice coil actuation coefficient, $A_l$. It is given by
\begin{equation}
A_l(f_1) = A_f(f_2) \, \frac{S_f}{S_l} \, \frac{R(f_1)}{R(f_2)} \, \frac{r(f_1)}{r(f_2)} = A_f(f_2)\,\frac{S_f}{S_l}\,\frac{R_l}{R_f}\,\frac{r_l}{r_f}
\label{Aetm1}
\end{equation}
where subscripts $l$ and $f$ identify terms that are associated with, or measured at the frequency of, the length and frequency excitations, $f_1$ and $f_2$, respectively; $R_l$ and $R_f$ are the amplitudes of the modulations in the arm locking readout signal; $S_l$ is the amplitude in the ETM length excitation monitor point signal (see figure~\ref{FreqShifter}); $r_l$ and $r_f$ are the closed-loop responses of the single-arm cavity locking readout signal to length fluctuations. The $r_l/r_f$ ratio allows propagation of the calibration from the VCO excitation frequency to the voice coil excitation frequency. This ratio is determined either by sequential excitation at frequencies $f_1$ and $f_2$ with a characterized length actuator or by a model of the single-arm closed-loop servo based on the measured response.

\begin{figure}
 \begin{flushright}
     \includegraphics[width=0.8\textwidth]{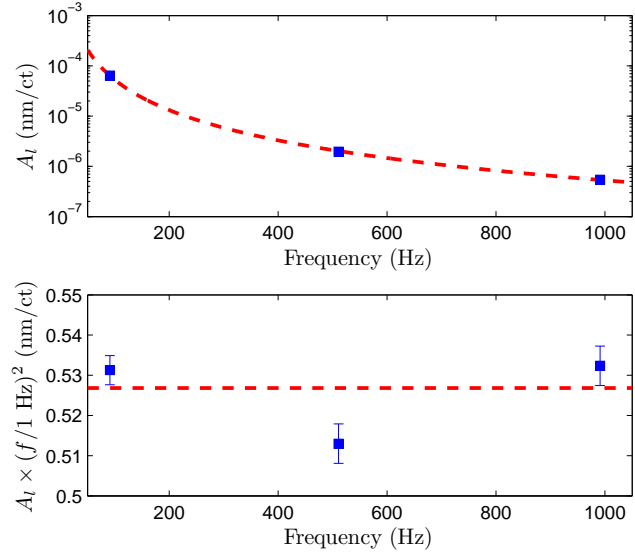}
 \end{flushright}
 \caption{ETM voice coil actuation coefficients measured using the frequency modulation technique for the $x$-arm of the Hanford 4-km-long interferometer. The dashed lines are a weighted least-squares fit with a $f^{-2}$ functional form. In the lower panel, the expected functional dependence is removed by multiplying by the square of the measurement frequency. The error bars represent the estimated $\pm 1 \sigma$ uncertainties.}
 \label{vcoResults}
\end{figure}

The frequency modulation calibration method was applied to the $x$-arm of the Hanford 4 km interferometer at three widely separated frequencies within the most sensitive region of the LIGO detection band, 91, 511, and 991 Hz. The data for these measurements were recorded by driving the frequency modulations at all three frequencies simultaneously, each with an associated length modulation separated by 0.1 Hz (six excitations total). For the typical four-minute integration times, the signal-to-noise ratio (SNR) of the voice coil actuation peak in the single-arm readout signal is limited to about 30 to avoid saturation of the actuation electronics. The SNR of the frequency modulation actuation peak is about 1000. However, the SNR of the peak in the frequency modulation monitor signal is only 10 because the monitor signal is also the control signal for the mode cleaner locking servo, and thus has an elevated noise floor. The calibration results are plotted in figure~\ref{vcoResults}. The dashed lines denote a weighted least-squares fit to the data with the expected $f^{-2}$ force-to-length functional form. The error bars show the estimated $\pm 1 \sigma$ uncertainties of approximately 0.8\%, as described in section~\ref{Errors}.

The accuracy of this method was investigated by comparing calibration results with those derived from the free-swinging Michelson and photon calibrator methods described in section~\ref{Intro}~\cite{CalCompare}. The standard deviation of the mean values (over all frequencies) of the actuation coefficients derived using these three techniques for the four ETMs at the Hanford observatory (two interferometers) was less than 2.5\%. This suggests that the accuracy of the applied frequency modulation method is within this range.

\section{Estimate of uncertainties}\label{Errors}
Expanding equation~\ref{Aetm1}, the calibrated voice coil actuation function can be written as,
\begin{equation}
\label{Aetm}
A_l = \left[\frac{-C(f)L}{\nu}\right] \, 2\mathcal{K}(f) \, H_{mc}(f) \left[\frac{S_f}{S_l} \, \frac{R_l}{R_f}\right] \frac{r_l}{r_f}.
\end{equation}
The uncertainty in $A_l$ is estimated by calculating partial derivatives of equation~\ref{Aetm} with respect to variables that have significant uncertainties and summing in quadrature. These estimates are summarized in table~\ref{tab:VcoErrors} and discussed below.
\begin{table}
	\caption{\label{tab:VcoErrors}Summary of the significant relative uncertainties contributing to the overall relative uncertainty for the frequency modulation voice coil calibration technique.}
	\begin{indented}
	  \item[]\begin{tabular}{@{}ll}
	    \br
			Variable & $1\sigma$ uncertainty \\
			\mr
			$S_f$ actuation function, $\mathcal{K}$ & 0.1 \% (Statistical) \\
			Mode cleaner filtering, $H_{mc}$ & 0.05 \% ($f_0\sim3$\%) \\
			Signal ratio, $(S_f R_l)/(S_l R_f)$ ($N \simeq 35$) & 0.8 \% (Typical) \\
			Control loop response, $r_l/r_f$ & 0.05 \% (Typical) \\
			{\bf Estimated overall uncertainty} & {\bf 0.8 \%} \\
			\br
		\end{tabular}
	\end{indented}
\end{table}

The first term in square brackets has negligible uncertainty because both $L$, and therefore $C(f)$, and $\nu$ are known with high accuracy. The relative uncertainty in arm cavity length is approximately $10^{-4}$~\% and the relative uncertainty in laser frequency is approximately 0.01\%.

The statistical uncertainty in the $S_f$ calibration function, $\mathcal{K}$, is determined using two methods. First, repeated measurements are made at a single frequency; second, measurements are made at multiple frequencies over a span from 90 Hz to 2 kHz (see figure~\ref{calVCOdata}). Both methods yield a standard error in the calibration function of approximately 0.1\%, originating from uncertainty in measuring the sideband-to-carrier power ratio. We have not included estimates of potential sources of systematic error associated with measuring the sideband-to-carrier power ratio using a spectrum analyzer. We expect them to be small because the power levels typically differ by only 30~dB and the frequencies ($\approx$80~MHz) are separated by less than 2~kHz.

The results of repeated measurements of the mode cleaner pole frequency vary by as much as 3\%. However, the contribution of this variation to the overall uncertainty in the actuation coefficient is reduced by a factor of $(f_0^2/f^{2} + 1)^{-1}$ due to the partial derivative of equation~\ref{Aetm} with respect to $f_0$. Thus, the contribution to the uncertainty in $A_l$ due to uncertainty in the mode cleaner pole frequency is about 0.05\% at 1~kHz and even smaller at lower frequencies.

The last term in square brackets includes the frequency and length excitation amplitudes measured at the monitor points and the measured amplitudes in the single-arm readout signal. For typical measurements, we use 4 minute-long Fourier transforms and 35 averages. This reduces the combined standard error for this term to 0.8\%, dominated by the uncertainties in measurement of $S_f$ and $R_l$. The contribution from $S_f$ is large because the monitor point for the frequency excitation is downstream of the summation point for the mode cleaner locking servo. $R_l$ contributes significantly due to the small excitation amplitudes used in order to avoid saturation of the ETM actuation electronics.

For the 0.1 Hz frequency separation used in these measurements, the estimated $r_l/r_f$ ratio differs from unity by less than 0.05\%, significantly below measurement statistical variations. This was confirmed experimentally by repeating calibration measurements with the length and frequency modulation excitation frequencies interchanged.

Adding all of these relative uncertainties in quadrature, we estimate the typical fractional $1\sigma$ uncertainty in the calibration of the ETM voice coil actuation coefficient to be approximately 0.8\%. With longer integration times and more averaging of the measured signals, the overall estimated uncertainty could be further reduced.

\section{Conclusions}
We have described a new technique for calibrating the test mass displacement actuators of the LIGO interferometers that uses frequency modulation of the injected laser light to create an effective length modulation fiducial. We have also described the method employed to measure the amplitude of the applied frequency modulation and therefore the induced effective length modulation. Procedures used to improve the overall estimated test mass voice coil calibration precision to less than 1\% ($1\sigma$) have been discussed.

The test mass actuation coefficients determined using this technique are consistent with those derived using two distinctly different methods, the free-swinging Michelson and the photon calibrator. Unlike both of these methods, the frequency modulation technique does not exert additional forces directly on a test mass. Measurements and finite-element modeling have shown that elastic deformation of the test masses caused by these actuation forces can induce large errors in actuator calibration, especially for actuation frequencies above 1~kHz~\cite{LIGOPcal,HildEffect,TMdeform}.

For the frequency modulation method, we induce effective arm length displacements of approximately $10^{-13}$ m which are much smaller than the displacements used for the free-swinging Michelson method ($\sim$$10^{-8}$ m) but much larger than those used for the photon calibrator method ($\sim$$10^{-17}$ m). In contrast to the free-swinging Michelson method that requires multiple sequential measurements, frequency modulation enables a single-step actuator calibration. However, it uses a single-arm configuration rather than the full science mode configuration in which searches for gravitational waves are performed and the photon calibrator method is applied.

Recent improvements in the single-arm feedback control loop have reduced noise levels. This should enable increased calibration precision with shorter integration times using the frequency modulation method. Furthermore, measurements carried out over the past year have shown the long-term stability of the voice coil actuation functions is better than 1\%~\cite{LongTermCal}. Thus, repeated frequency modulation calibration measurements should yield a highly accurate measurement of the overall statistical variation for this voice coil calibration method.

Increasing measurement precision and accuracy and applying several disparate calibration methods has improved our understanding of systematic errors and increased our confidence in test mass actuator calibration results~\cite{CalCompare}. However, optimizing the scientific reach of future gravitational wave searches will require further improvements in detector calibration accuracy and precision~\cite{ReqRespErrors}. We expect that the frequency modulation method will continue to play a role in these efforts.

\ack We gratefully acknowledge enlightening discussions with R. Adhikari, K. Kawabe, M. Rakhmanov, P. Schwinberg, D. Sigg, and the LIGO Calibration team. We also thank the National Science Foundation for support under grant PHY-0555406. LIGO was constructed by the California Institute of Technology and Massachusetts Institute of Technology with funding from the National Science Foundation and operates under cooperative agreement PHY-0107417. This paper has LIGO Document Number LIGO-P0900259.

\section*{Appendix. Calibration of the frequency actuator}\label{Afcal}
\appendix
\setcounter{section}{1}
The AOM at the heart of the laser frequency shifter (see figure~\ref{FreqShifter}) uses the first-order beam that is Bragg-diffracted from the acoustic wave driven by the radio-frequency (RF) signal from the AOM driver.  Energy and momentum conservation dictate that the frequency of the laser light in this beam is up-shifted (or down-shifted) by the frequency of the acoustic wave~\cite{AOMref} which is dictated by the frequency of the signal from the AOM driver.  The frequency shifter bandwidth is greater than 1 MHz, so for the range of modulation frequencies used (up to 2~kHz) the frequency modulation of the diffracted light is given by the frequency modulation of the RF signal driving the AOM.

To characterize the AOM driver, a phase-locked loop (PLL) is used to lock its output frequency to a frequency standard. This minimizes frequency drifts enabling precise measurement of the amplitudes of the RF carrier ($\sim$80~MHz) and modulation sidebands using an RF spectrum analyzer. The unity gain frequency of this PLL is approximately 400~Hz. The AOM driver input monitor signal is calibrated by injecting a sinusoidal frequency excitation, measuring the amplitude of the sinusoidal signal at the $S_f$ monitor point, and using an RF spectrum analyzer (Agilent 4395A) to measure the ratio of the power in one of the induced first-order frequency modulation sidebands with respect to the carrier in the AOM driver output signal.

The time-varying electric field of the frequency-modulated AOM driver output signal can be expressed as
\begin{equation}
E(t) = E_0 e^{i(\omega t + \phi(t))}
\end{equation}
where $E_0$ is the amplitude of the sinusoidally varying electric field, $\omega$ is angular frequency of the RF carrier, and
\begin{equation}
\phi(t) = \int_0^t{\Delta \omega \cos (2 \pi f \tau) {\rm d} \tau}=\Gamma \sin (2 \pi f t),
\end{equation}
with the modulation index, $\Gamma$, given by $\Gamma=\Delta\omega/(2\pi f)$. The frequency-modulated field can be decomposed into a carrier and a series of frequency-shifted sideband fields by writing it as an infinite series of Bessel functions of the first kind, $J_n$, as
\begin{equation}
E(t) = E_0 e^{i \omega t} \sum_{n = - \infty}^{\infty} J_n (\Gamma) e^{2 i \pi n f t}.
\end{equation}
The ratio of the power in one of the first-order sidebands with respect to the power in the carrier is then given by $P_1/P_0 = J_1^2(\Gamma)/J_0^2(\Gamma)$. With the measured carrier and sideband powers, this expression yields $\Gamma$ and therefore $\Delta\omega$, the amplitude of the frequency modulation of the RF signal driving the AOM. As discussed above, this is equivalent to $\pi\Delta\nu$ where $\Delta\nu$ is the amplitude of the laser frequency modulation resulting from double-passing the AOM as shown in figure~\ref{FreqShifter}.

The $S_f$ calibration function, $\mathcal{K}\equiv\Delta\omega/(2\pi S_f)=\Delta\nu/(2S_f)$, measured at frequencies between 90~Hz and 2~kHz is plotted in the upper panel of figure~\ref{calVCOdata}. To interpolate to other modulation frequencies and to assess the frequency response of the VCO, we also measure the transfer functions of the analog electronics denoted by blocks A and B in figure~\ref{FreqShifter}, $H_A(f)$ and $H_B(f)$. The VCO actuation coefficient, $\alpha$, which we expect to be frequency-independent, is determined by a least-squares fit using pole-zero approximations of the measured electronics transfer functions, with $\alpha$ as the only free parameter. Thus, $\mathcal{K}(f) = \alpha H_B(f)/H_A(f)$. The lower panel of figure~\ref{calVCOdata} shows the normalized deviations between the measurements and the fit and indicates that $\alpha = 90.62 \pm 0.09$~Hz/count is frequency independent, as expected.


\section*{References}
\begin{thebibliography}{99}
   \bibitem{OcalPaper} Landry~M for the LIGO Scientific Collaboration 2005 {\it \CQG} {\bf 22} S985--94
   \bibitem{LIGOPcal} Goetz~E, \etal 2009 {\it \CQG} {\bf 26} 245011 (13pp)
   \bibitem{GEOCal} Hewitson~M for the LIGO Scientific Collaboration 2007 {\it \CQG} {\bf 24} S445--55
   \bibitem{VirgoCal} Rolland~L 2009 to be submitted to {\it Amaldi 8 Conf. Proc.}
   \bibitem{S5paper} Kissel~J, \etal 2010 Calibration of the LIGO instruments for the fifth science run {\it LIGO Technical document} P0900120 (https://dcc.ligo.org/)
   \bibitem{GEOPcal} Mossavi~K, Hewitson M, Hild S, Seifert F, Weiland U, Smith J R, L\"uck H, Grote H, Willke B and Danzmann K 2005 {\it \PL}A {\bf 353} 1--3
   \bibitem{VirgoPcal} Rolland~L, Marion~F and Mours~B 2008 {\it Virgo doc.} VIR-053A-08 (https://pub3.ego-gw.it)
   \bibitem{CalCompare} Goetz~E, \etal 2010 {\it \CQG} {\bf 27} 084024 (11pp)
	 \bibitem{DynamResonance} Rakhmanov~M, Savage~Jr~R~L, Reitze~D~H and Tanner~D~B 2002 {\it \PL} A {\bf 305} 239--44
   \bibitem{HildEffect} Hild~S, \etal 2007 {\it \CQG} {\bf 24} 5681--8
   \bibitem{TMdeform} Afrin~Badhan~M, Landry~M, Savage~R and Willems~P 2009 Analyzing elastic deformation of test masses in LIGO {\it LIGO doc.} T0900401 (https://dcc.ligo.org)
	\bibitem{MalikThesis} Rakhmanov~M 2000, PhD Thesis California Institute of Technology
  \bibitem{ReqRespErrors} Lindblom~L 2009, {\it \PR}D {\bf 80} 042005-1--7
  \bibitem{LongTermCal} Savage Jr R L, \etal 2010 eLIGO Photon Calibrator Investigation: Long-Term Stability of DARM Actuation G1000254 (https://dcc.ligo.org)
  \bibitem{AOMref} Yariv~A 1985 {\it Optical Electronics} 3rd Ed. (New York: CBS College Publishing) p~388
\end{thebibliography}
\end{document}